# Beating the reaction limits of biosensor sensitivity with dynamic tracking of single binding events

Derin Sevenler[a,b], Jacob Trueb[c] and M. Selim Ünlü[a]
  a. Department of Electrical and Computer Engineering, Boston University, Boston, MA, 02215
  b. Center for Engineering in Medicine, Massachusetts General Hospital, Harvard Medical School, Boston, MA 02114
  c. Department of Mechanical Engineering, Boston University, Boston, MA, 02215

The clinical need for ultra-sensitive molecular analysis has motivated the development of several endpoint assay technologies capable of single molecule readout. These endpoint assays are now primarily limited by the affinity and specificity of the molecular recognition agents for the analyte of interest. In contrast, a kinetic assay with single molecule readout could distinguish between low abundance, high affinity (specific analyte) and high abundance, low affinity (nonspecific background) binding by measuring the duration of individual binding events at equilibrium. Here we describe such a kinetic assay, in which individual binding events are detected and monitored during sample incubation. This method uses plasmonic gold nanorods and interferometric reflectance imaging to detect thousands of individual binding events across a multiplex solid phase sensor with a large area approaching that of leading bead-based endpoint assay technologies. A dynamic tracking procedure is used to measure the duration of each event. From this, the total rates of binding and de-binding as well as the distribution of binding event durations are determined. We observe a limit of detection of 15 femtomolar for a proof-of-concept synthetic DNA analyte in a 12-plex assay format.

Several of the promises of precision medicine rely on ultra-sensitive molecular diagnostic technologies. Liquid biopsies of circulating genomic, transcriptomic, or proteomic biomarkers of cancer promise earlier detection and treatment, as well as improved guidance of targeted therapies in treating minimum residual disease (1, 2). Similarly, sensitive and specific molecular diagnostic tests for infectious pathogens are vital for the identification and management of pre- or asymptomatic individuals (3, 4). Likewise, panel assays of circulating biomarkers could soon improve the accuracy of diagnosis of injuries such as acute liver failure or traumatic brain injury (5, 6).

These pressing clinical needs have motivated the development of a variety of ultra-sensitive assay technologies, culminating in technologies capable of single molecule detection. A unifying characteristic of essentially all of these assay technologies is that they employ molecular recognition agents (or 'capture probes') such as antibodies, nanobodies, peptides, oligonucleotides, aptamers, or other agents that bind specifically to the molecule of interest. Single molecule detection technologies commonly then use droplet emulsions (7) or microwell arrays (8) to isolate and then enumerate the precise number of analyte molecules bound to the capture probes.

In terms of signal transduction, it is clear that single molecule detection is 'as good as it gets' (9). But, transducing the amount of captured analyte is only one half of the picture—the analyte must be captured in the first place. Even with single molecule detection, assay performance is still limited by the affinity of the capture probes. This causes sensitivity and specificity to vary widely between different probe-analyte pairs. For example, it is now relatively routine to quantify some molecular species (e.g. genomic DNA) with single-copy sensitivity and precision (10) while the detection limits of other analytes (e.g. microRNA) are many orders of magnitude worse (11–13). Probe affinity can also vary between samples. Variations in extensive properties of the sample such as pH and ion content change the free energy of binding, and variable amounts of nonspecific background binding complicates quantitation further.

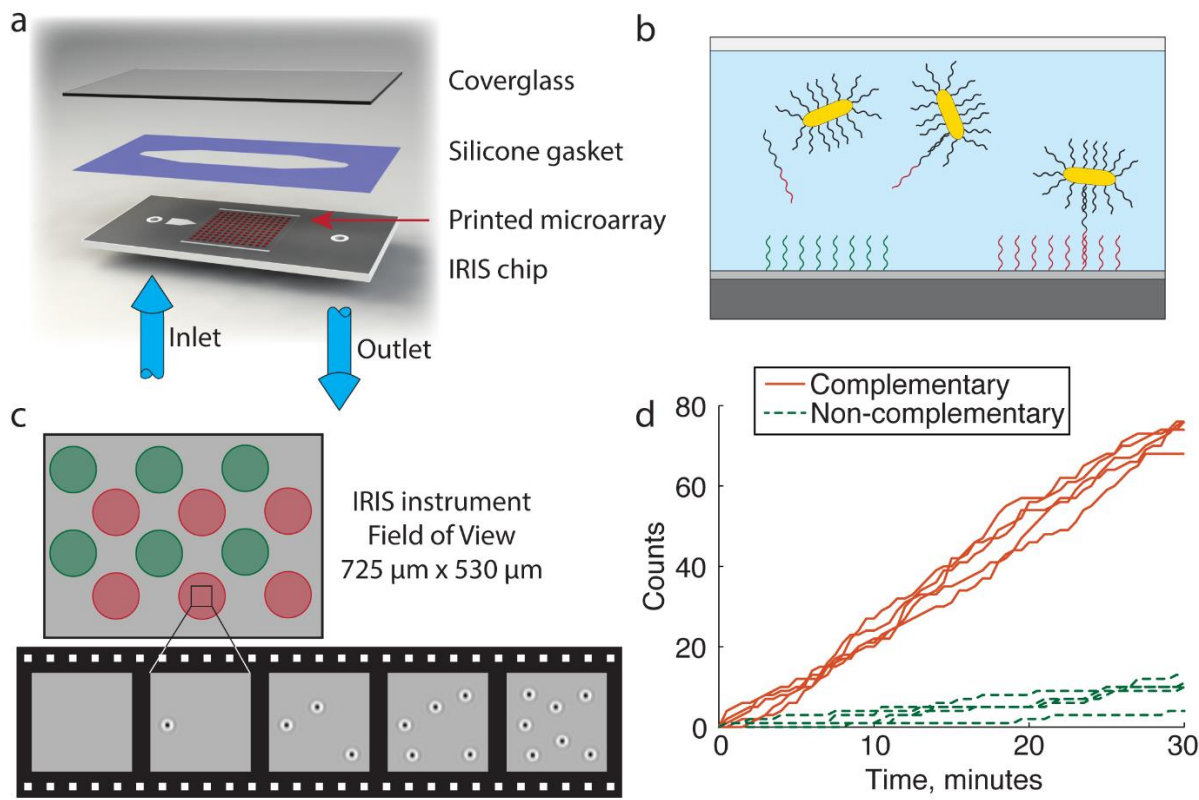

**Figure 1:** Dynamic measurements of single binding events across a large sensor surface. (a) Rendered image of IRIS chip perfusion chamber for dynamic measurements of molecular interactions. A DNA microarray is printed on the IRIS chip. Then, the chamber is formed by layering a patterned adhesive gasket and anti-reflection coated coverglass viewing window. The IRIS chip has two through-holes for sample perfusion. The entire disposable costs about $5 USD. (b) Nucleic acid assay with IRIS. DNA-conjugated gold nanorods (GNRs) are pre-incubated with the sample solution and hybridize with complementary nucleic acids. The mixture is flowed over the chip. Nucleic acid analytes strands hybridize to complementary GNRs and tether them to adjacent-complementary DNA spots. (c) Schematic of dynamic detection of single GNRs with IRIS. Images are simulated. GNRs on the chip surface are observed as diffraction-limited spots and automatically detected using purpose-built software. (d) Plots of total nanoparticle binding to six complementary (red) and six non-complementary (green) DNA spots over time, as measured with dynamic tracking, for a single experiment where the target concentration is 300 fM.

Current leading single molecule detection technologies rely on signal amplification reactions. These are endpoint assays: the probe molecules are incubated with the sample for a set amount of time, after which the reaction is halted so that amplification and detection can be performed. What is measured is the amount of bound analyte at the instant the incubation is halted.

In contrast to endpoint assays, kinetic assays directly measure probe-analyte interactions during course of the incubation. Kinetic assays collect more information than endpoint assays: they can measure not just concentration but also molecular affinity *via* rates of association and dissociation. This could allow for inter-sample variations in probe affinity or nonspecific binding to be identified and mitigated without additional tests. For low concentrations of analyte, kinetic assays could also be capable of distinguishing low-abundance specific binding from a larger background of nonspecific binding, or even measuring analytes below the so-called 'critical concentration' at which there is fewer than one analyte molecule bound at equilibrium—a feat impossible for endpoint assays (14).

However, single molecule kinetic measurements are technically demanding: without amplification reactions, specific binding events are more difficult to discern against a background of nonspecific interactions. Indeed, an exquisitely sensitive transduction mechanism is required to directly detect single binding events at all. A range of scientific apparatuses have been developed to investigate single molecule binding kinetics. However, none of these techniques are useful for ultra-sensitive clinical assays because the sensors are too small. To investigate nanoscale phenomena, these devices are themselves nano- or micro-scale: their active sensors are the size of single nanoparticles or nanowires (15–17), or else they require high magnification & high numerical aperture optics with a small field of view (0.001-0.01 mm$^2$) (18–20). This is problematic because small sensors only have space for a small number of capture probes. Maximizing the number of probes is vital for ultra-sensitivity: at low concentrations, the amount of captured analyte at equilibrium is proportional to the number of probe molecules. Assay technologies therefore use large sensor areas packed with capture probes. For example, the SiMoA technology interrogates approximately 25,000 beads each 2.7 µm in diameter, corresponding to a total sensor area of 0.57 mm$^2$ (12).

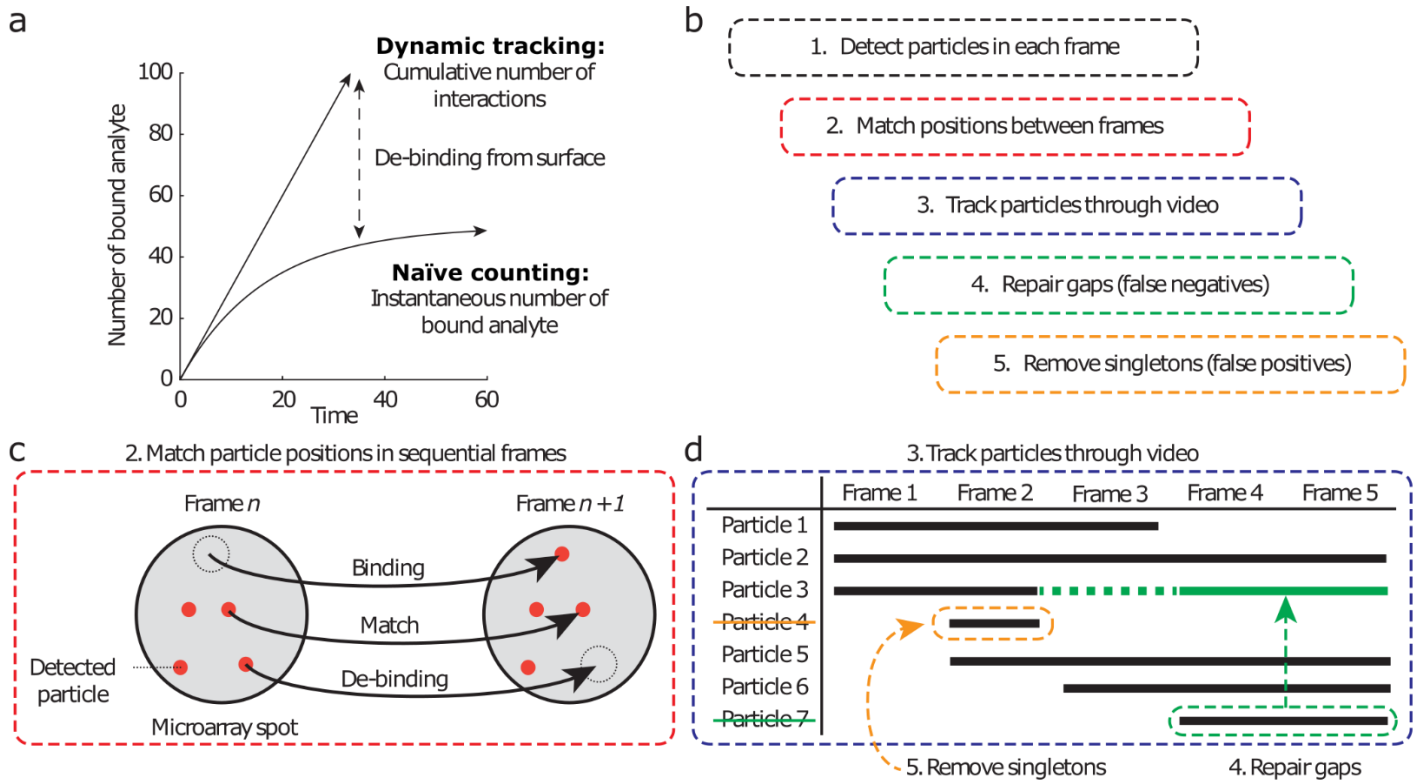

**Figure 2:** (a) Dynamic tracking improves sensitivity when the de-binding rate is nonzero. Consider a reaction limited Langmuir model where most binding sites remain open at equilibrium (e.g. Figure 1d). Equilibrium is reached when the rate of new analyte binding is balanced by analyte de-binding from the sensor. Dynamic tracking measures the cumulative number of analyte binding events, rather than the instantaneous number of bound analyte over time (Naïve counting). (b-d) Diagram of the multi-step dynamic tracking algorithm used to track individual GNR labels in IRIS images (described in text).

Here, we describe a kinetic assay that measures the duration of individual binding events over time on a large sensor surface with a low magnification objective, while retaining the advantages of kinetic analysis such as discrimination between specific and nonspecific events based on event duration. In this study, we used a 20x, 0.45 NA objective and a 1.1" format camera which yielded a sensor area of 0.38 mm$^2$, comparable with that of ultra-sensitive endpoint methods. (this area could be further increased several-fold with different optical instrumentation and stage scanning).

**Results**

**Detection of individual binding events across a large field of view.** We recently described the development of a 'digital microarray' assay technology which rapidly enumerates individual captured molecules across hundreds of microarray spots (21). This technology uses probe-conjugated gold nanorods (GNRs) as molecular labels, and an interferometric reflectance imaging sensor (IRIS) to rapidly detect individual GNRs with a 10x microscope objective. The large field of view enabled a similar throughput to commercial fluorescence readers while enhancing the limit of detection and dynamic range by a factor of approximately 10,000. The detection reaction is based on the bio-barcode assay developed by others (22) and is compatible with a range of analytes.

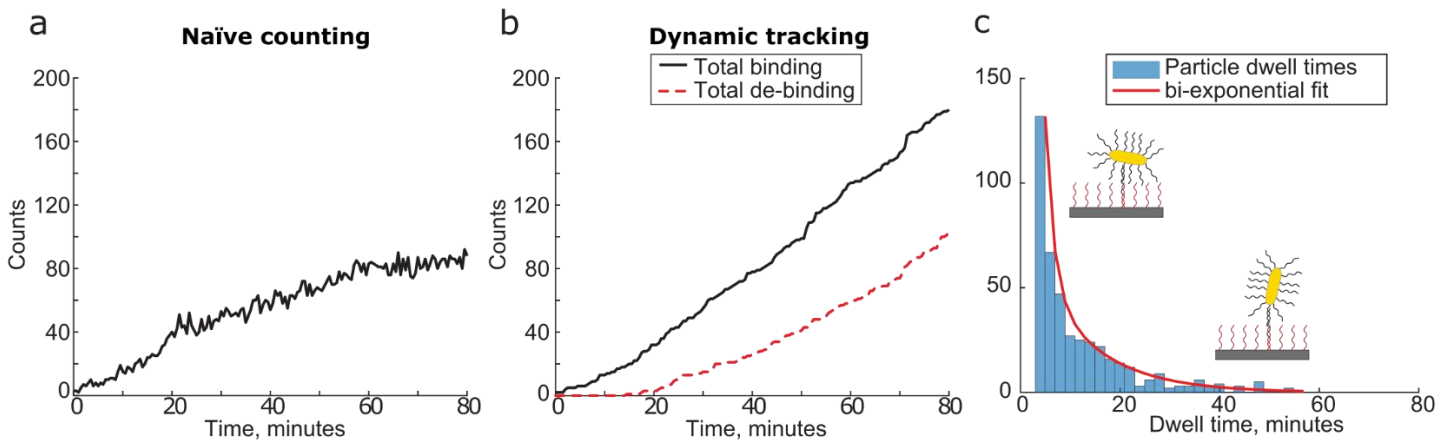

**Figure 3:** Experimental comparison of dynamic tracking with naïve counting. (a) Instantaneous number of GNRs binding to complementary and non-complementary control spots over time. The analyte concentration is 316 fM and the GNR concentration is 20 pM. The sensor reaches equilibrium within 90 minutes. (b) Total number of GNR binding and de-binding for the complementary spot in (a), as measured with dynamic tracking. The rate of total GNR binding is constant, and equilibrium is reached when the rates of binding and de-binding are equal. (c) Histogram of complementary spot dwell times across all experiments, with a biexponential fit. This distribution may be caused by differences in affinity between side-immobilized and end-immobilized GNRs.

We adapted the IRIS digital microarray platform for dynamic measurements by designing a perfusion chamber that consists of an IRIS chip, a patterned silicone gasket, and an antireflection-coated coverglass window (Figure 1a). Two holes for the chamber inlet and outlet are drilled in the IRIS chips by wafer-scale laser micromachining. The assembly is held by a custom clamp fixture that makes fluidic connections to the inlet and outlet on the bottom of the chip (Figure S1).

To demonstrate dynamic detection of single molecules, synthetic 'target' ssDNA oligonucleotides were pre-incubated with target-complementary DNA conjugated gold nanorods 25 nm x 70 nm (GNRs) for ninety minutes. The concentration of GNR labels was kept constant for all experiments at 20 pM while the concentration of the target varied from 10 pM down to 10 fM. After pre-incubation, the mixture was perfused over IRIS chips with DNA microarrays of target-complementary and noncomplementary probes, and GNR-labeled targets hybridized to the complementary spots (Figure 1b).

Images were acquired every 30 seconds with the IRIS instrument during perfusion. GNRs on the IRIS chip were visible as faint diffraction-limited blobs in the images, which were detected in each frame independently using custom software (Figure 1c). The rates of GNR binding were then measured using a dynamic tracking algorithm, described in the following section (Figure 1d).

IRIS detects individual GNRs based on light scattering. Since water has a higher refractive index than air, the polarizability and scattering cross section of the GNRs was reduced compared with dry chips. Additionally, the image suffered from spherical aberrations caused by the air-coverglass interface. GNR visibility was restored by switching from a 10x, 0.3 NA objective to a 20x, 0.45 NA coverglass corrected objective. The resulting field of view of 0.38 mm$^2$ (725 µm by 530 µm) could comfortably accommodate 12 microarray spots each approximately 80 µm in diameter.

**Dynamic tracking of binding events over time.** Under sufficiently high flow rates, the initial rate of binding of analyte is proportional to the bulk analyte concentration. One may estimate the analyte concentration by plotting the number of bound GNRs over time, and measuring the initial slope (Naïve counting, Figure 2a). However, this approach has several problems. The first is the fundamental issue related to finite probe affinity mentioned earlier. Ultra-low analyte concentrations will reach equilibrium with very few (or even less than one) bound molecules. In those cases, the initial slope will not be measurable even with perfect error-free readout. The second issue is that some unbound GNRs are detected in each frame as they transiently diffuse through the detection volume. These transient particles result in false positives that increase the overall noise floor of the sensor.

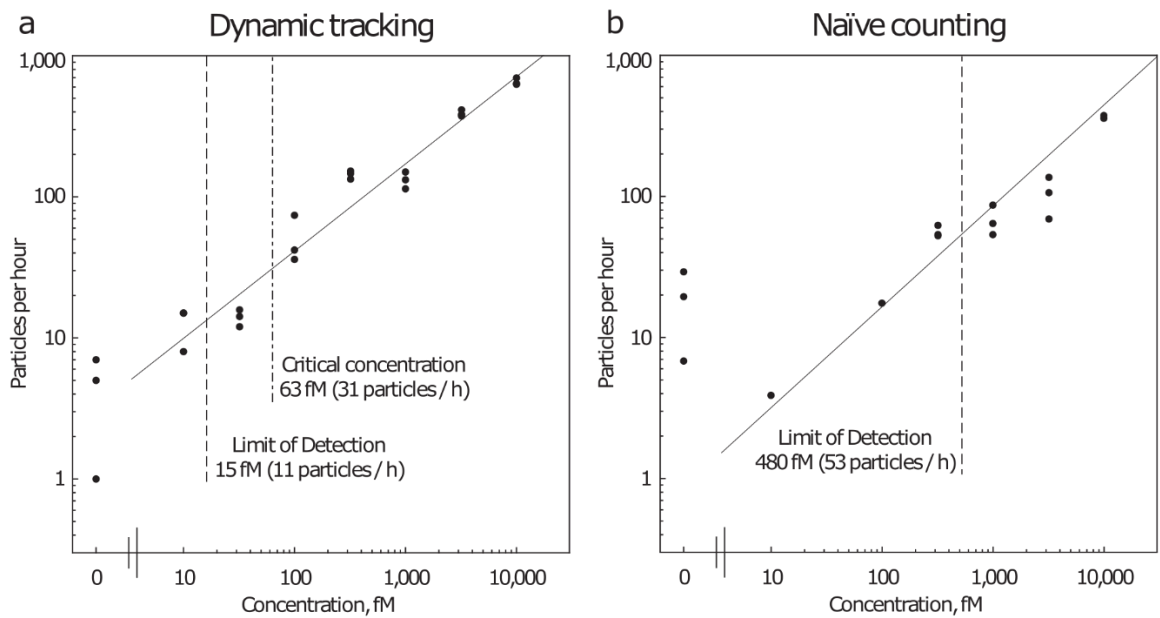

**Figure 4:** Rate of particle binding versus analyte concentration for the same experiments, using (a) dynamic tracking or (b) naïve counting. The solid lines are linear regressions on the log-log data. The dashed lines indicate the Limit of Detection (LOD) for both techniques, as well as the critical concentration at which there is fewer than one analyte bound at equilibrium. Naïve counting had a negative overall count for some spots in the 10, 32 and 100 femtomolar chips which cannot be plotted. Dynamic tracking improved the LOD by about 30-fold compared with naïve counting and resulted in a highly linear response over a much wider range of analyte concentrations, as expected. Notably, the dynamic tracking LOD is four-fold lower than the critical concentration.

To address these issues, we developed a post-processing algorithm that uses the spatial positions of particles to track them individually over the course of the experiment (Figure 2b). First, particles are detected in each video frame independently of other frames, and their positions in the image are recorded. Second, these positions are compared with those of particles in the next frame of the video (Figure 2c). Particles in the same position in both frames are 'matched', indicating that they are in fact the same particle. Particle matching includes a clustering algorithm that is robust to small translations between frames. For a video with $N$ timepoints, this results in $N - 1$ lists of matches. Third, these lists are compiled into a single master catalog which tracks the contiguous series of frames in which each particle was observed (Figure 2d). This is essentially a table which lists when each particle bound, where it bound, and when (if ever) it de-bound from the surface. Finally, this catalog is filtered to reduce false positives and false negatives. False negatives occur when a particle is mistakenly not detected in a single frame but was detected in the same position in previous and subsequent frames. This erroneously results in two entries in the catalog. These gaps are repaired by identifying whether the binding of each particle corresponds to the exact same place as the de-binding of another particle two frames prior, and then merging the two catalog entries (e.g. particles #3 and #7, Figure 2d). False positives are caused by particles visible in just one frame, and are simply removed (e.g. particle #4, Figure 2d).

This catalog can then be used to plot the rate of new binding events. For low analyte concentrations, most of the available binding sites will remain empty at equilibrium and the rate of new binding events will be constant and proportional to the bulk concentration. At an analyte concentration of 316 femtomolar for example, the sensor reached equilibrium with about eighty bound GNRs after one hour (Figure 3a). As predicted, the rate of target-hybridized GNR binding was constant over time, even after equilibrium was reached ('Total binding', Figure 3b). Note that the total number of binding and de-binding events are both cumulative measurements and are therefore monotonically increasing over time.

We compared the measured rates of binding with the predicted rates of transport of analyte-bound GNRs to the spots (Supporting Information). As expected, we found that the measured rate of binding was well below the theoretical upper limit predicted by mass transport, due to the finite binding rate $k_{on}$.

**Measurements of nanorod dwell times.** The duration or 'dwell time' of each binding event may be measured using dynamic tracking. Note that the dwell time can only be measured for particles which did de-bind before the end of the experiment. Taken together, these dwell times allow the off rate $k_{off}$ of the GNRs to be determined (Figure 3c). We found that our experimental results were best explained by a bi-exponential fit of the form $N(t) = A_1 e^{-k_1 t} + A_2 e^{-k_2 t}$. A histogram of dwell times across all complementary spots was generated for each experiment and fitted independently (Figure S3). Experiments that were either too brief or contained too few binding events to have meaningful statistics were discarded. Of the remaining experiments, the fitting parameters were mostly consistent and did not trend with analyte concentration (Figure S3). The average values were $k_1 = 0.53$ min$^{-1}$ and $k_2 = 0.082$ min$^{-1}$ and $A_1/A_2 = 25$, corresponding with primary and secondary dissociation time constants $\tau_1 = 1.9$ min and $\tau_2 = 12$ min. We were initially surprised that a bi-exponential fit explained the distribution much better than a single exponential, and suspected that this may be due to GNRs tethered by two or more analyte molecules. However, we do not believe this was the case because the relative weights between the two terms $A_1/A_2$ was similar across a large range of concentrations and did not decrease for lower concentrations. Since the total GNR concentration was kept constant at 20 pM, the relative number of GNRs with two bound analytes (*versus* one bound analyte) should decrease with decreasing analyte concentration.

We speculate that the bi-exponential distribution in dwell times was caused by the geometry of the GNR labels. The binding energy is likely greater if the rod is tethered to the surface by one end, rather than by the middle, for two reasons. First, there is electrostatic repulsion between the DNA-functionalized GNRs and the DNA-coated chip surface. A side-tethered GNR is likely be constrained to bring a larger area closer to the surface. Second, the end-tethered GNR has a larger number of conformational degrees of freedom (three rotational DOF) than a side-tethered one (one rotational DOF), resulting in a greater entropic penalty to binding. Since the rods are functionalized uniformly across their surfaces with label sequences, GNRs will be more likely to capture analyte to their sides rather than ends during pre-incubation since it represents a greater percentage of their surface area.

Note that we cannot differentiate between dissociation of the analyte from the surface probe and from the GNR label. In the model system used here, both surface-analyte and analyte-label duplexes are 25 base pairs long and similar GC content. Therefore, their affinities should be similar, and they should be responsible for GNR dissociation at roughly similar rates.

**Detection below the critical concentration.** A standard curve was measured by performing identical experiments with a range of analyte concentrations between 10 fM and 10 pM. GNR binding to three complementary spots were analyzed using both dynamic tracking and naïve counting (Figure 4). A linear least-squares regression was performed on the log-transformed data, and the limit of detection (LOD) was calculated as three standard deviations above the mean signal from the blank sample. The LOD for dynamic tracking of 15 fM surpassed that of naïve counting by over a factor of 30 (480 fM). Additionally, dynamic tracking had a much more linear response over the tested dynamic range (3 orders of magnitude).

Notably, dynamic tracking had a limit of detection nearly four-fold lower than the critical concentration of this assay. Equivalently: dynamic tracking was able to detect the presence of the analyte even when the average duration of binding events was shorter than the time between binding events. At the critical concentration, equilibrium is reached with one bound analyte molecule on average. This dissociation rate is simply the weighted average of the two dissociation constants found earlier: $\bar{k}_{off} = (A_1 k_1 + A_2 k_2)/A_1 A_2 = 0.52$ min$^{-1} = 31$ particles per hour. The critical concentration was found by taking intercept of the dynamic tracking regression line with this binding rate: $c^* = 63$ fM. As expected, sensitivity was limited only by nonspecific binding rather than insufficient probes even in this 12-plex assay format.

**Discussion**

A range of endpoint assay technologies have been developed that have single molecule readout. For these assays, the limiting factor is instead the affinity and specificity of the molecular recognition agents. The number of capture probes (for example, the number of functionalized beads) can almost always be increased until sensitivity is limited by nonspecific binding rather than insufficient numbers of probes (23), but further improvement must come through careful optimization of washes and reactions. Protocol optimization are particularly challenging for multiplexed test development, since the optimal wash conditions (duration, ion content, surfactants, pH, and so on) are likely different for different probe-analyte complexes.

In response to these limitations, we have introduced a kinetic assay technology which measures the duration of individual binding events across a large sensor area. In this work we distinguished specific and nonspecific binding events without even a single wash step, which could have been used to further improve specificity. Notably, kinetic analysis alleviates the need for a globally optimal wash protocol and therefore makes multiplexed tests straightforward.

'Solid-phase' surface sensors are sometimes criticized for having poor mass transport kinetics, as compared with bead-based assays. We alleviated this effect by using a high flow rate, which makes the depletion layer very thin (Supplementary Information). As a result, equilibrium was reached even for the lowest concentrations in under two hours. Of course, peristaltic pumping and re-circulation can be used to re-circulate small sample volumes for longer experiments.

Somewhat unexpected bi-exponential behavior suggested two different conformations of immobilized GNRs, each with different binding free energy: end-tethered and side-tethered. This variability complicates the probe-analyte affinity measurement and could be problematic for heterogeneous samples. This could be further tested and perhaps mitigated by preferentially functionalizing just the ends of GNRs (24, 25).

Because these proof-of-concept experiments were conducted with synthetic analytes and pure buffer solutions, direct comparisons with more mature assay technologies cannot yet be made. Future verification of this technique is envisioned in multiplexed quantification of messenger- and microRNAs towards a range of potential clinical applications.

## Materials and Methods

Additional materials and methods are available in the Supporting Information at the end of this document.

**Perfusion chamber assembly:** No. 1 coverslips 25.4 mm by 12.7 mm with a broadband antireflection coating on one side were purchased from Abrisa Technologies (Torrance, CA). Custom patterned silicon gaskets with were purchased from Grace Biolabs (Bend, OR). Silicone gaskets were 25.4 mm by 12.7 mm, 0.15 mm thickness, with pressure-sensitive adhesive on one side. In preparation, gaskets were adhered to the non-coated side of the coverglass and stored with protective tape in place.

The perfusion chamber was assembled by aligning the gasket-window assembly to the IRIS chip, loading it into the clamp fixture, removing the protective tape and engaging the clamp to form seals between the chip and the gasket as well as with the sample inlet and outlet. The volume of the chamber was approximately 8 µL.

**IRIS digital microarray instrument for dynamic detection:** The operating principle of IRIS is thoroughly described elsewhere (26). Briefly, the IRIS instrument consists of a reflectance microscope with a single high-powered LED for illumination (M660L4 LED with FB650-10 bandpass filter, Thorlabs) and a monochrome machine vision camera (Grasshopper GS3-U3-123S6M-C, Point Grey Research). The digital microarray implementation of IRIS is optimized for rapidly detecting individual gold nanorods based on their anisotropic light scattering properties. The design, optimization, and implementation of the optical system has been described in detail elsewhere (21). For dry IRIS chips, this system can detect single gold nanorods with a 10x, 0.3 NA objective. For dynamic experiments the system was entirely the same except that a 20x, 0.45 NA coverslip-corrected air immersion objective (Nikon CFI S Plan Fluor ELWD 20x) was used. The higher light collection efficiency compensated for the decreased intensity of GNR light scattering due to immersion of the rods in water, and the collar allowed correction of spherical aberrations from the coverslip-air interface.

**Assay protocol:** The assay protocol was identical for all experiments, except that the concentration of the target analyte was changed. First, the DNA-GNR conjugates and target DNA oligos were pre-mixed in a 'hybridization buffer' consisting of 10 mM phosphate pH 7.4, 600 mM Na+, 0.1% Tween-20, and 1 mM EDTA. 100 µL of GNRs stored at 200 pM were mixed with 900 µL of hybridization buffer containing the target DNAs. The final GNR concentration was 20 pM for all experiments. The mixture was vortexed briefly and sonicated for 10 seconds before storing at room temperature in a microcentrifuge tube. After 90 minutes, the sample was aspirated with a disposable 1 mL needle-tipped syringe. The needle tip was removed, and the syringe was connected to a Luer fitting on the end of the inlet tube. The outlet waste tube was left in conical vial. The syringe was mounted in a syringe pump and the sample was dispensed at 10 µL/min for up to 90 minutes. The instrument was re-focused as soon as the liquid sample filled the chamber. Video acquisition began 1-3 minutes after the sample first contacted the chip surface.

**Image acquisition:** Image acquisition was automated using the Micro-manager (27) microscope control software with custom scripts. Scripts have been made freely available online at www.github.com/derinsevenler/IRIS-API. Timepoints were taken every 30 seconds at each timepoint, a z-stack of nine images was acquired with a step size of two microns (i.e., a span of 16 microns). At each z-position, four images were acquired and averaged pixel-wise before saving to reduce shot noise.

**Image processing and particle detection:** The video data from each experiment consisted of an image hyperstack of 180 (t) x 9 (z) x 12.4 MP (x,y). A 86 µm by 86 µm (500 by 500 pixels) region of the video was cropped around each microarray spot. GNRs in each region and timepoint were detected independently. The particle detection method described here is a refinement of methods described earlier (21, 28), and has three steps: preprocessing, key point detection and key point filtering. First, a sparse pseudo-median filter is applied to each frame of the z-stack (made available online by others at http://imagejdocu.tudor.lu/doku.php?id=plugin:filter:fast_filters:start) to estimate the image background. True median filtering is effective for removing punctate features but computationally expensive for larger kernels. We found the sparse pseudo-median algorithm preferable due to its speed. Next, the normalized intensity image was calculated by pixelwise division of the original frame from the background. Finally, the 'normalized intensity range' (NIR) image was measured by projecting the maximum difference (i.e., max - min) at each pixel of the normalized intensity stack. Although not every GNR is visible in every normalized intensity image in the stack, each particle is clearly visible in the resulting NIR image.

Key points in the NIR images were detected by applying a global threshold to binarize the image. Blobs in the binary image (i.e., regions brighter than the threshold) were enumerated and then filtered based on size and shape. Specifically, a minimum area, maximum area, and minimum area-perimeter ratio were specified. The detection threshold and key point filtering parameters were manually selected and then kept constant for all experiments.

**Acknowledgements**

DS is grateful to Fulya Ekiz Kanik, Joeseph Greene, and Abdul Bhuiya for feedback on the image acquisition and analysis software and perfusion chamber design; to Celalettin Yurdakul for help designing the chip lithography pattern; and to Professor Ahmad Khalil for fruitful discussions about these results.

**References**

1. Cohen JD, et al. (2017) Combined circulating tumor DNA and protein biomarker-based liquid biopsy for the earlier detection of pancreatic cancers. *PNAS* 114(38):10202–10207.

2. Alix-Panabières C, Pantel K (2016) Clinical Applications of Circulating Tumor Cells and Circulating Tumor DNA as Liquid Biopsy. *Cancer Discov*. doi:10.1158/2159-8290.CD-15-1483.

3. Shurtleff AC, Whitehouse CA, Ward MD, Cazares LH, Bavari S (2015) Pre-symptomatic diagnosis and treatment of filovirus diseases. *Front Microbiol* 6. doi:10.3389/fmicb.2015.00108.

4. Alsdurf H, Hill PC, Matteelli A, Getahun H, Menzies D (2016) The cascade of care in diagnosis and treatment of latent tuberculosis infection: a systematic review and meta-analysis. *The Lancet Infectious Diseases* 16(11):1269–1278.

5. Antoine DJ, et al. (2013) Mechanistic biomarkers provide early and sensitive detection of acetaminophen-induced acute liver injury at first presentation to hospital. *Hepatology* 58(2):777–787.

6. Papa L, et al. (2016) Time Course and Diagnostic Accuracy of Glial and Neuronal Blood Biomarkers GFAP and UCH-L1 in a Large Cohort of Trauma Patients With and Without Mild Traumatic Brain Injury. *JAMA Neurol* 73(5):551–560.

7. Vogelstein B, Kinzler KW (1999) Digital PCR. *PNAS* 96(16):9236–9241.

8. Cohen L, Walt DR (2017) Single-Molecule Arrays for Protein and Nucleic Acid Analysis. *Annual Review of Analytical Chemistry* 10(1):345–363.

9. Walt DR (2012) Optical Methods for Single Molecule Detection and Analysis. *Anal Chem* 85(3):1258–1263.

10. Schoepp NG, et al. (2017) Rapid pathogen-specific phenotypic antibiotic susceptibility testing using digital LAMP quantification in clinical samples. *Science Translational Medicine* 9(410):eaal3693.

11. Rissin DM, et al. (2017) Polymerase-free measurement of microRNA-122 with single base specificity using single molecule arrays: Detection of drug-induced liver injury. *PLOS ONE* 12(7):e0179669.

12. Cohen L, Hartman MR, Amardey-Wellington A, Walt DR (2017) Digital direct detection of microRNAs using single molecule arrays. *Nucleic Acids Res* 45(14):e137–e137.

13. Wu D, Katilius E, Olivas E, Dumont Milutinovic M, Walt DR (2016) Incorporation of Slow Off-Rate Modified Aptamers Reagents in Single Molecule Array Assays for Cytokine Detection with Ultrahigh Sensitivity. *Anal Chem* 88(17):8385–8389.

14. Squires TM, Messinger RJ, Manalis SR (2008) Making it stick: convection, reaction and diffusion in surface-based biosensors. *Nat Biotech* 26(4):417–426.

15. Baaske MD, Foreman MR, Vollmer F (2014) Single-molecule nucleic acid interactions monitored on a label-free microcavity biosensor platform. *Nat Nano* 9(11):933–939.

16. Besteman K, Lee J-O, Wiertz FGM, Heering HA, Dekker C (2003) Enzyme-Coated Carbon Nanotubes as Single-Molecule Biosensors. *Nano Lett* 3(6):727–730.

17. Sorgenfrei S, et al. (2011) Label-free single-molecule detection of DNA-hybridization kinetics with a carbon nanotube field-effect transistor. *Nat Nano* 6(2):126–132.

18. Halpern AR, Wood JB, Wang Y, Corn RM (2014) Single-Nanoparticle Near-Infrared Surface Plasmon Resonance Microscopy for Real-Time Measurements of DNA Hybridization Adsorption. *ACS Nano* 8(1):1022–1030.

19. Piliarik M, Sandoghdar V (2014) Direct optical sensing of single unlabelled proteins and super-resolution imaging of their binding sites. *Nat Commun* 5:4495.

20. Lamichhane R, Solem A, Black W, Rueda D (2010) Single-molecule FRET of protein–nucleic acid and protein–protein complexes: Surface passivation and immobilization. *Methods* 52(2):192–200.

21. Sevenler D, Daaboul GG, Ekiz Kanik F, Ünlü NL, Ünlü MS (2018) Digital Microarrays: Single-Molecule Readout with Interferometric Detection of Plasmonic Nanorod Labels. *ACS Nano* 12(6):5880–5887.

22. Hill HD, Mirkin CA (2006) The bio-barcode assay for the detection of protein and nucleic acid targets using DTT-induced ligand exchange. *Nat Protoc* 1(1):324–336.

23. Yelleswarapu VR, Jeong H-H, Yadavali S, Issadore D (2017) Ultra-high throughput detection (1 million droplets per second) of fluorescent droplets using a cell phone camera and time domain encoded optofluidics. *Lab Chip* 17(6):1083–1094.

24. Caswell KK, Wilson JN, Bunz UHF, Murphy CJ (2003) Preferential End-to-End Assembly of Gold Nanorods by Biotin−Streptavidin Connectors. *J Am Chem Soc* 125(46):13914–13915.

25. Pramod P, Joseph STS, Thomas KG (2007) Preferential End Functionalization of Au Nanorods through Electrostatic Interactions. *J Am Chem Soc* 129(21):6712–6713.

26. Sevenler D, Avci O, Ünlü MS (2017) Quantitative interferometric reflectance imaging for the detection and measurement of biological nanoparticles. *Biomed Opt Express, BOE* 8(6):2976–2989.

27. Edelstein AD, et al. (2014) Advanced methods of microscope control using μManager software. *J Biol Methods* 1(2). doi:10.14440/jbm.2014.36.

28. Trueb JT, Avci O, Sevenler D, Connor JH, Ünlü MS (2017) Robust Visualization and Discrimination of Nanoparticles by Interferometric Imaging. *IEEE Journal of Selected Topics in Quantum Electronics* 23(2):1–10.

# Supporting Information

## Supplemental Materials and Methods

**IRIS chip fabrication:** 150 mm polished silicon wafers with nominally 110 nm of thermally grown oxide were purchased from Silicon Valley Technologies (SVM). This film thickness was selected since it maximizes the visibility of nanorods with a longitudinal surface plasmon resonance peak at 650 nm (6). SVM also performed photolithography and oxide etching to pattern the chips with identifying features and dicing lines. Wafers were protected with a layer of photoresist before shipping to Patomac Laser (Baltimore, MD) for through-hole drilling. Finally, the wafers were diced into 25.4 mm by 12.7 mm rectangular chips and stripped of photoresist at Boston University. Chips were inspected under the microscope to ensure cleanliness and stored in a sealed container.

**Microarray printing:** A 100 mm disposable plastic petri dish containing 10 mL of MCP-4 coating solution (Lucidant Inc, Sunnyvale CA) was prepared following manufacturer's instructions. IRIS chips were exposed to either pure oxygen or air plasma for five minutes to activate the surface with silanol groups. Chips were immediately submerged in the coating solution after plasma treatment and placed on an orbital shaker at room temperature for 30 minutes, during which the polymer covalently bonded to the glass *via* trimethoxysilane moieties. Chips were then washed thoroughly in DI water and dried in a vacuum oven at 80C for 15 minutes. Chips were inspected under the microscope to ensure a clean, even coating and stored in a desiccator at room temperature for up to two weeks.

For microarray spotting, two vials containing one each of the amine-functionalized ssDNA surface probes (Table 1) were prepared with a final concentration of 25 μM DNA, 150 mM phosphate pH 8.5. The high pH is required to facilitate reaction of the primary amines on the surface probes with N-hydroxysuccinimide moieties on the MCP-4 polymer. Single droplets of approximately 200 pL of the spotting solutions were printed onto the chip with a S3 Spotter (Scienion Inc, Berlin Germany) in a controlled humid chamber with 70% relative humidity. An interlocking pattern was used to reduce unused space on the array, such that each of the two conditions was printed in a square pattern with a pitch of 250 μm and offset by 125 μm horizontally and vertically from the other condition. Images taken by the spotter during printing were inspected to ensure droplets did not run together. The resulting spots were approximately 70-80 μm in diameter, with gaps of about 100 μm between them (Figure 1c). Chips were left in the spotter overnight to maximize immobilization and washed thoroughly the next morning in saline-sodium

## Estimation of the rate of mass transport

We estimated the rate of transport of target-captured gold nanorods to the sensor surface, using the notation described by Squires and Manalis, 2008 (1). The reader is encouraged to consult this text for derivations and detailed descriptions of the dimensionless numbers used in the following brief analysis.

The flow cell cross section is a rectangular channel, width $W_c = 4$ mm and height $H = 150$ µm. Each microarray spot was approximately 75 µm in diameter, which we approximate as an equal-area square $L = W_s = 66.5$ µm. The sensor to channel height ratio was therefore $\lambda = \frac{H}{L} = 2.25$. The volumetric flow rate $Q$ was 10 µl/min. The diffusion coefficient of the nanorods was estimated based on kinetic theory (below) to be 6.6 µm2/s.

If we assume that all targets hybridize to nanorods before flow begins, we can predict the rate of target-hybridized nanorod interactions with each microarray spot based on mass transport. The channel Peclet number $Pe_H = \frac{Q}{W_c D} \approx 6300$ is large, suggesting that convection is much faster than diffusion and the depletion layer is small compared to the chamber height. The shear Peclet number $Pe_S = 6\lambda^2 Pe_H = 1.9 \times 10^5$ is also very large, indicating the depletion layer is very thin compared to the sensor itself, i.e., the sensor is not really depleting the sample at all. Only target-hybridized nanorods that happen to be very close to the surface (within about 1 µm) have a chance of being captured.

The flux of target-hybridized nanorods to the surface is therefore linearly proportional to the bulk concentration $c_0$ as $J_D \approx 0.81 \, Pe_S^{\frac{1}{3}} W_s D c_0$. A target concentration of 100 fM would correspond with a flux of about 1.8 particles per second per spot, or roughly 6,500 particles per hour per spot. The observed rate of interactions is somewhat less than 1% of this theoretical upper limit (Figure 4). This difference may be partially due to incomplete capture of analyte by GNRs during pre-incubation. However, we believe that this difference is mostly due to the lower 'on-rate' $k_{on}$ of analyte-bound GNRs than the analyte alone. Consider for example that the pose of the GNR must be correct for the analyte to be presented downwards towards the DNA probes. The presence of the GNR also decreases the favorability of binding due to electrostatic repulsion between it and the DNA microarray spot.

## Diffusion coefficient of the nanorods

The total translational diffusion coefficient of the DNA-coated nanorods was approximated based on Stokes-Einstein equation from kinetic theory. For a spherical particle of radius $r$, the diffusion coefficient is given by

$$D_{sphere} = \frac{k_B T}{6\pi \eta r}$$

where $r$ is the particle radius, $\eta$ the solvent dynamic viscosity and $k_B T$ the particle's average thermal energy. For a spheroid, the diffusion coefficient must be adjusted by a 'friction factor' first described by Perrin (2) that depends on the particle aspect ratio:

$$D_P = \frac{D_{sphere}}{f_P}$$

Here $D_{sphere}$ is the diffusion coefficient of a sphere of equivalent volume. The friction factor for a prolate spheroid is calculated from the particle aspect ratio $p = a/b$, where a and b are the spheroid major and minor axes respectively ($a > b$), by (3)

$$f_P = \frac{\sqrt{p^2 - 1}}{p^{\frac{1}{3}} \ln(p + \sqrt{p^2 - 1})}$$

Although gold nanorods tend to be rod-shaped rather than prolate, the error is small when $p < 10$ (3). The bare gold nanorods in these experiments were nominally 25 nm by 71 nm, and functionalized with a dense coat of ssDNA 30 nucleotides long. Immobilized DNA in the coat are thought to extend outwards due to electrostatic repulsion from each other (4, 5). We somewhat arbitrarily estimated the thickness of this coat as 10 nm based on the known relation for rigid double-stranded dsDNA of 0.34 nm per base pair. For the purposes of estimating their diffusion coefficients therefore, rods were estimated as 45 nm by 90 nm prolate spheroids with $p = 2$, $f_p = 1.05$. A sphere of equivalent volume would have a radius of $r = 31$ nm, giving $D_{sphere} = 6.9$ µm2/s. and $D_P = 6.6$ µm2/s.

## Considerations for improving sensor area and dynamic range

The particle detection method used here detects bound GNRs at each timepoint independently of previous timepoints. This contrasts with methods in which binding events are detected differentially, by subtracting each frame or timepoint from the previous one. This technique is therefore much more robust to small displacements caused by vibration and unlocks the possibility of increasing the sensor area (and therefore sensitivity and/or multiplexing) by scanning multiple fields of view. A bidirectional repeatability of < 5 µm is easily achievable using modern linear positioners, and well within the tolerance of this method even without image registration.

Perhaps the main limiting factor is that imaging elsewhere would reduce the frequency of timepoints. However, other considerations include the effect of transient local heating due to light absorption by the nanoparticles. When the illumination beam is kept stationary, the system is in pseudo-equilibrium and the effect is consistent across all conditions and timepoints. Switching the beam from one region to another would introduce temporal variations in heating which could affect the thermodynamics of binding.

The upper limit of quantification is set by the concentration of gold nanorods (20 pM), which must be in sufficient excess of the analyte. The dynamic range of this technique could consequently be improved by increasing the GNR concentration. Each experiment consumed about 20 femtomoles of GNRs, which had a reagent cost of roughly $1.20. It should be noted however that increasing the GNR concentration could have a range of adverse effects on assay performance by increasing the rate of nonspecific binding and transient unbound particles in the images.

# Supplemental Figures

**Video SV1:** Timelapse of IRIS images during an incubation with 100 fM target, with Dynamic Tracking results. The shown image is a 30 µm by 30 µm region cropped from one complementary spot, representing 0.2% of the full Field of View. Black blobs are the diffraction limited images of individual gold nanorods as detected with a 20x objective. Each tracked particle in the catalog (Figure 2d) is circled with a unique color.

| Name (number of nucleotides) | Sequence |
|---|---|
| GNR label sequence (30) | <u>TTCTCGATCAGCTTCTCTTTTACGC</u> AAAAA -ThioMC6 |
| Target (50) | <u>GCGTAAAAGAGAAGCTGATCGAGAA</u>**AGCAGGCGCGTGGTACAGCTACAAA** |
| Complementary surface probe (37) | AmMc6- GGGAAAAAAGGG**TTTGTAGCTGTACCACGCGCCTGCT** |
| Non-complementary surface probe (37) | AmMc6- GGGAAAAAAGGGCTTCGGCAATACCGCCCATACCGGC |
| Stabilizer sequences | CCCTTTTTTCCC |

**Table 1:** DNA oligonucleotide sequences used in this study. "AmMC6" indicates 5'-end amino functionalization for linking to the IRIS chip, while "ThioMC6" indicates 3'-end thiol functionalization for GNR conjugation, as provided by Integrated DNA Technologies. The GNR label sequence and surface probe each have a 25-nucleotide sequence complementary to one half of the target molecule. The surface probes also have a double-stranded 'stabilized' region between the surface and the target-complement region, to accelerate target hybridization.

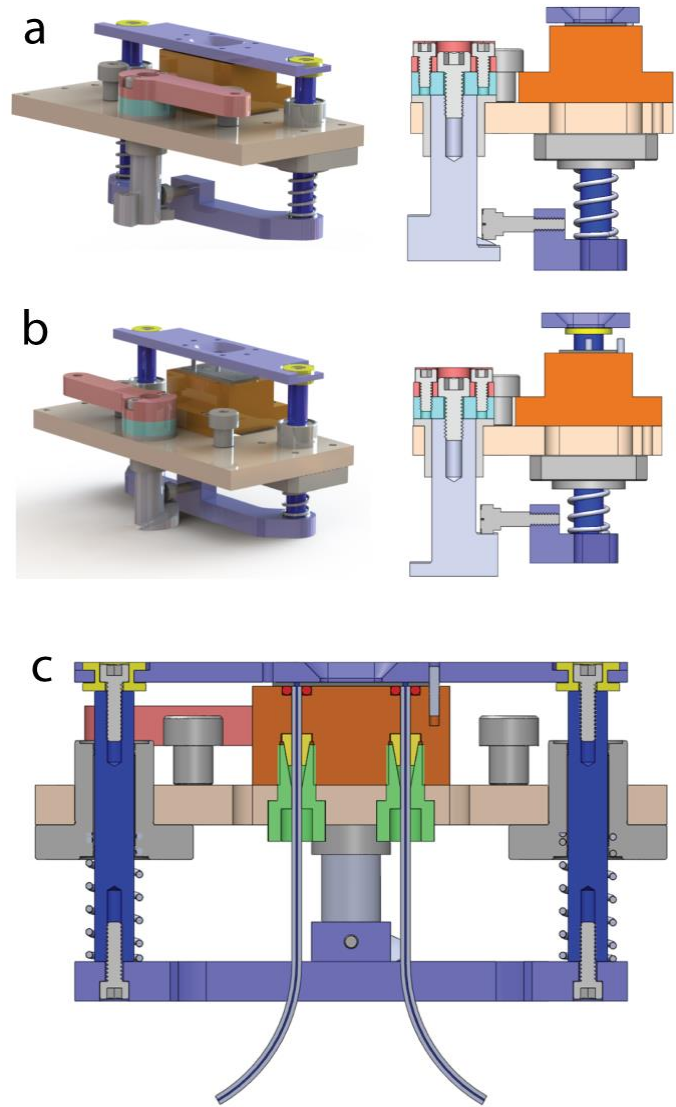

**Figure S1:** Renderings and drawings of perfusion chamber clamping fixture in (a) closed and (b) open position. Components are different colors in the rendering for visualization. Clamping force is applied to the top surface of the perfusion chamber by a spring-loaded removeable crossbar containing a square viewing window in its center. The clamping fixture can be opened by rotating a handle (pink) connected to a rotary cam mechanism, which compresses the springs and elevates the crossbar away from the perfusion chamber. The profile of the rotary cam decouples the spring force from the handle rotation while fully in the open position, allowing the operator to interact with the perfusion chamber without the need to continuously hold the clamping mechanism open. The side-cut view (c) shows fluid connections to the chip. PEEK tubing is held by finger-tight Upchurch fittings (green/yellow) and sealed by o-rings (red) against the chip inlet and outlet.

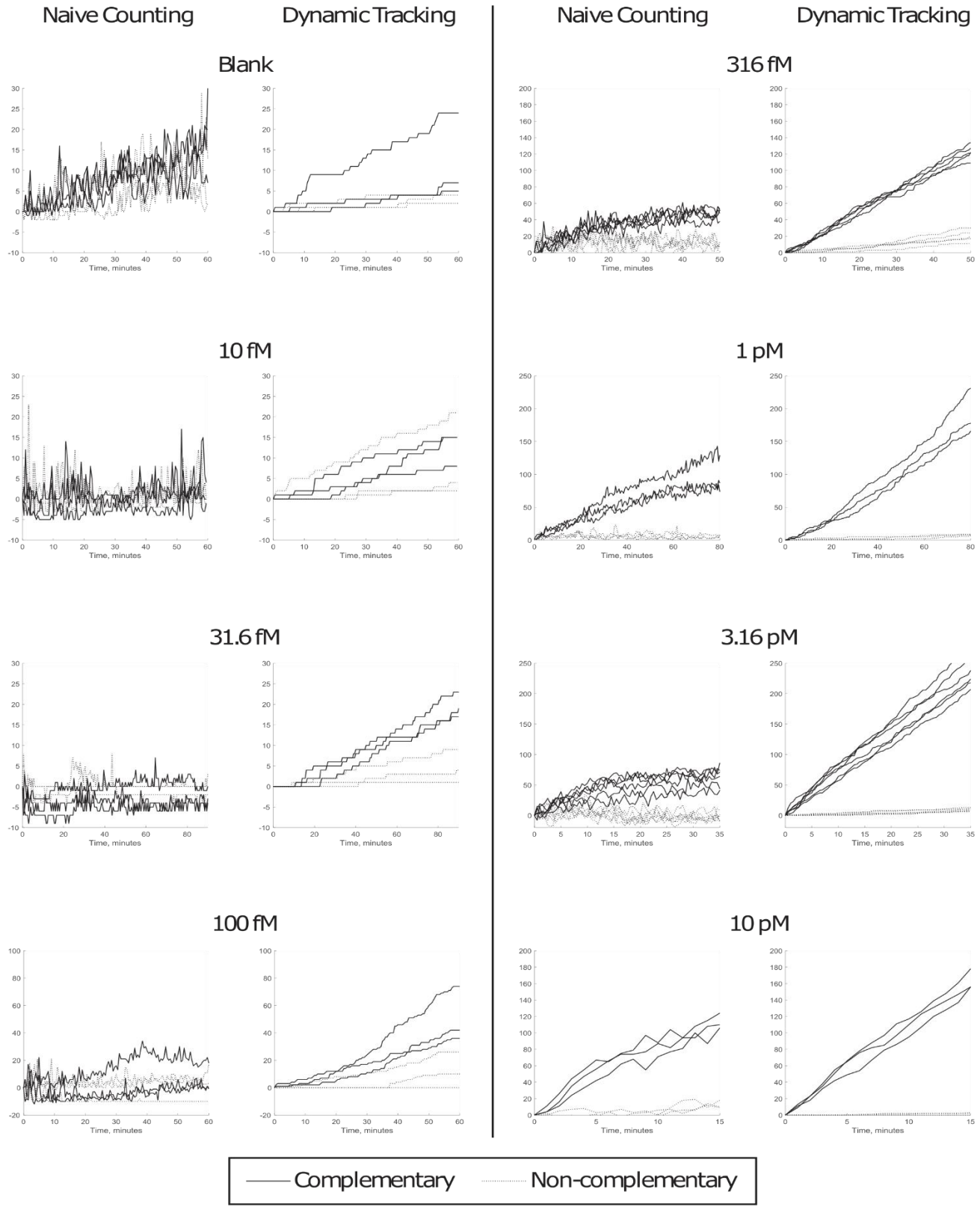

**Figure S2:** Traces of naïve counting and Dynamic Tracking for all spots from all experiments. Solid and dashed lines indicate complementary spots and non-complementary spots, respectively. While naïve counting plots the instantaneous number of binding events, Dynamic Tracking measures the cumulative number of binding events (compare with Figure 2a-b). Note that while the axes are identical for the naïve counting and Dynamic Tracking traces for a given experiment, both x- and y-axis scales vary between experiments. Three each of the complementary and non-complementary spots were analyzed for all experiments except 316 fM and 3.16 pM, for which six spots each were analyzed.

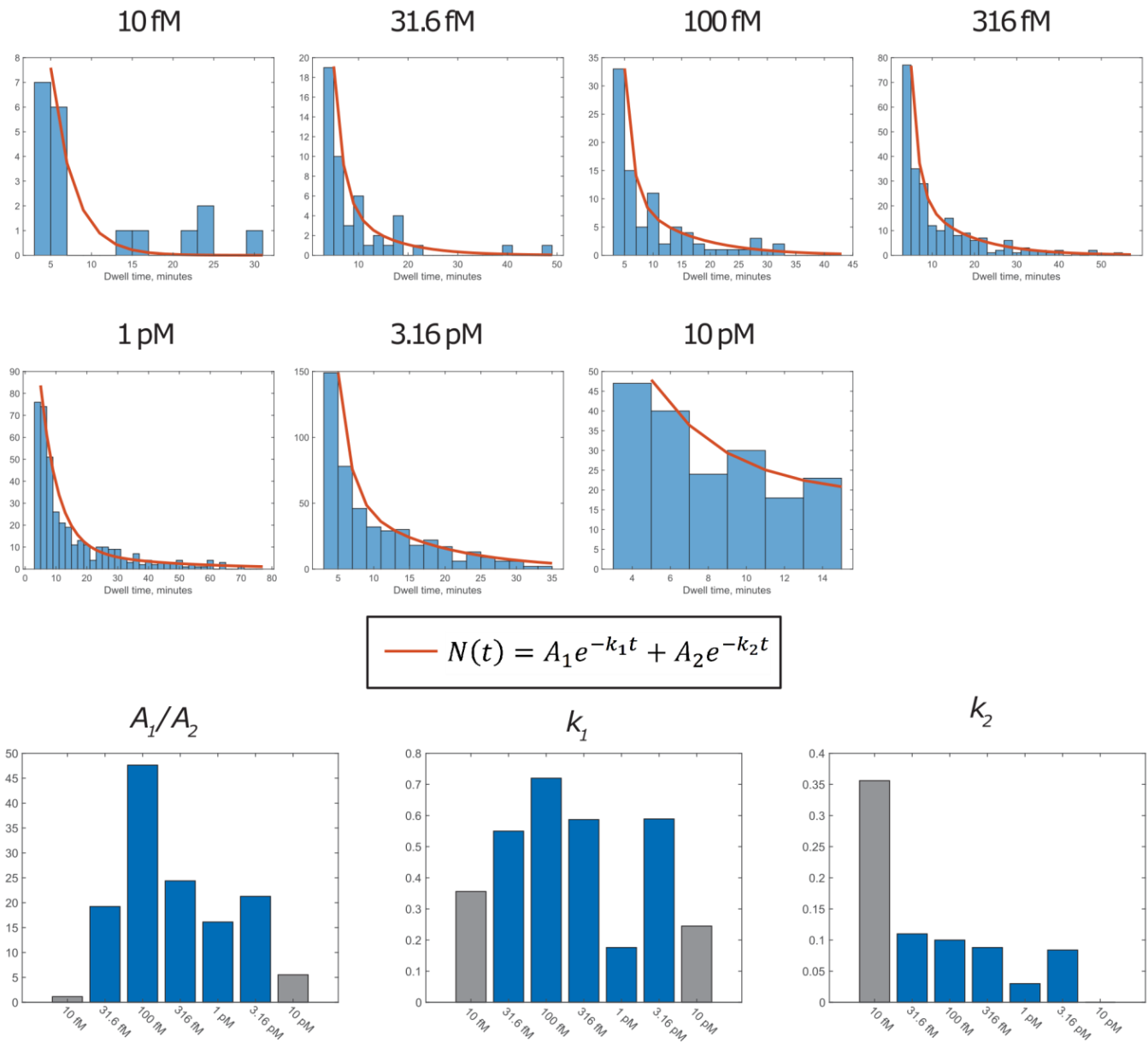

**Figure S3:** Histograms and bi-exponential fits of binding event durations ('dwell times') to complementary spots for all experiments. For each experiment, all binding events on the complementary spots were combined. Below, the fitting parameters for different experiments are plotted. Ignoring the lowest concentration (too few events, below the Limit of Detection) and highest concentration (duration too short) experiments, the fitting parameters are in rough agreement, and do not trend with increasing or decreasing analyte concentration.